\documentstyle[prl,aps,multicol,epsfig]{revtex}

\begin{document}

\renewcommand{\narrowtext}{\begin{multicols}{2}}
\renewcommand{\widetext}{\end{multicols}}
\newcommand{\lsim}   {\mathrel{\mathop{\kern 0pt \rlap
  {\raise.2ex\hbox{$<$}}}
  \lower.9ex\hbox{\kern-.190em $\sim$}}}
\newcommand{\gsim}   {\mathrel{\mathop{\kern 0pt \rlap
  {\raise.2ex\hbox{$>$}}}
  \lower.9ex\hbox{\kern-.190em $\sim$}}}
\def\be{\begin{equation}}
\def\ee{\end{equation}}
\def\ba{\begin{eqnarray}}
\def\ea{\end{eqnarray}}
\def\d{{\rm d}}
\def\i{{\rm i}}
\def\ap{\approx}
\def\L{{\mathcal L}}
\def\two{|l^+\nu\rangle_\pi}
\def\twot{|l^+\nu(t)\rangle_\pi}
\def\twoW{|l^+\nu\rangle_W}

\draft
\preprint{CERN--TH 2000-279}

\title{Two particle states, lepton mixing and oscillations}
\author{M. Kachelrie{\ss}$^1$, E. Resconi$^2$ and S. Sch\"onert$^{3,4}$}
\address{$^1$TH Division, CERN, CH--1211 Geneva 23} 
\address{$^2$Dipartimento di Fisica, Universit\'a di Genova, Via 
Dodecaneso, 33, I--16146 Genova }  
\address{$^3$\footnote{Permanent address}Max-Planck-Institut f\"ur
  Kernphysik, Saupfercheckweg 1, D--69117 Heidelberg}
\address{$^4$Research Center for Cosmic Neutrinos, Institute for
  Cosmic Ray Research, University of Tokyo}

\maketitle
\begin{abstract}
Discussions of lepton mixing and oscillations consider generally only
flavor oscillations of neutrinos and neglect the accompanying charged
leptons. In cases of experimental interest like
pion or nuclear beta decay an oscillation pattern is expected indeed
only for neutrinos if only one of the two produced particles is observed.
We argue that flavor oscillations
of neutrinos without detecting the accompanying lepton is
a peculiarity of the two-particle states $|l\nu\rangle$ produced in
pion or nuclear beta decay. Generally, an oscillation pattern is only
found if both particles are detected. 
We discuss in a pedagogical way how this distinction of the neutrinos  
arises, although on the level of the Lagrangian lepton mixing does not
single them out  against charged leptons. 
As examples, we discuss the
difference between the state $|l\nu\rangle$ produced by the decay 
of real $W$ boson and a $W$ originating from pion decay. 
\end{abstract}

\pacs{14.60Pq}

\narrowtext

\section{Introduction}

The Standard Model (SM) of particle physics has proven to be a firm basis
on which all our knowledge of this field rests since its construction
30 years ago \cite{SM}. Precision tests performed in the last decade
demonstrated in particular that it is also correct at the quantum
level. Novel phenomena such as neutrino masses or supersymmetric
particles, which cannot be accommodated within the SM, should not be
thought to contradict it, but rather to guide us as to new physics
beyond it.
Nowadays, this hunt for new phenomena is the main topic  
in particle physics. In contrast to the search for supersymmetry,
for which there is no positive signal so far%
\footnote{We should however mention that the annual modulation signal
   seen by the DAMA experiment can be consistently explained as the
   scattering of supersymmetric dark matter particles on their target
   nucleons~\cite{DAMA}. Also the experimental evidence for a light
   Higgs found at LEPII points towards low-energy
   supersymmetry~\cite{LEP}.}, there is mounting
experimental evidence for neutrino oscillations. On the one hand,
there are five solar neutrino experiments using different techniques
that see a deficit in the solar neutrino flux \cite{solarexp99}. Although this
deficit could have its origin in principle also in non-standard solar-
or nuclear physics, it can be shown that these explanations
are experimentally excluded \cite{SNP}. On the other hand, the case
for an oscillation solution to the atmospheric neutrino deficit has
become recently even stronger in the general perception. This is largely 
because both the zenith-angle distribution and the dependence of
the ratio $\nu_e/\nu_\mu$ as function of the ratio
(oscillation length)/(neutrino energy)
 found by the Superkamiokande collaboration
support the neutrino oscillation hypothesis~\cite{atmexp99}. 

Most aspects of neutrino oscillations have been discussed extensively
in the literature.
The usual derivation, presented e.g. in Ref.~\cite{dev}, of the
probability $P$ that a relativistic neutrino with momentum $p$ and
flavor $\alpha$ has the flavor $\beta$ after the time $t$,
\be   \label{P}
 P_{\nu_\alpha\to\nu_\beta}  = \left| \sum_l 
                        U_{\alpha l}^{(\nu)} U_{\beta l}^{(\nu)\ast}  
                        \exp\left(-\i E_l t\right) \right|^2 \,, 
\ee
uses only basic facts of quantum mechanics. 
Here, $E_l=\sqrt{m_l^2+p^2}$ is the energy of the neutrino mass
eigenstate $l$ and the matrix $U$ is commonly
chosen to represent the unitary transformation matrix between weak and
mass eigenstates of the neutrino.
In spite (or, perhaps, because) of the simplicity of its derivation,
Eq.~(\ref{P}) raises several conceptual questions.
The most prominent ones are if for the neutrino mass eigenstates a definite
energy or momentum should be used, under which conditions the use of
wave packets with smeared energy and/or momentum is necessary, the problem of
coherence, and the connection between the quantum mechanical treatment
and quantum field theory~\cite{cq}.

In this article we want to discuss a more basic question, namely why
mixing in the lepton sector reveals itself experimentally in flavor
oscillations of neutrinos but not of charged leptons. 
This question is motivated by the following simple fact: At the level of the
Lagrangian describing the charged-current interactions of leptons, 
\be   
 \L_{CC} = -
 \frac{g}{\sqrt{2}}\sum_{i,j} \bar l_{L,i} \gamma^\mu V_{ij} \nu_{L,j} 
                              W_\mu^- +{\rm h.c.} \,, 
\ee
the mixing $V=U^{({\rm ch})\dagger}U^{(\nu)}$ between charged leptons and
neutrinos has two different sources: It could be ascribed either
completely to mixing in the neutrino ($V=U^{(\nu)}$) or in the charged
lepton sector ($V=U^{({\rm ch})\dagger}$), or most probably to some
superposition of both. Physical results do not depend 
on the particular decomposition of $V$ in the Standard Model
with lepton mixing\footnote{We call SM with lepton mixing any model
  that allows non-zero neutrino masses but reproduces otherwise in the
  low-energy limit the SM. This means in particular that we do not
  consider sterile neutrinos.}. 
Thus, knowing only the charged-current Lagrangian, one would 
not expect any fundamental difference between neutrinos and charged
leptons in oscillation experiments. Main purpose of this article is
to clarify the exact reason for this difference.

Another formulation of this question is to ask for which conditions 
it is allowed to neglect the charged lepton produced together with 
the neutrino in a two-particle state, e.g., in the decay of a pion
or a real $W$ boson. We will show that there is a crucial difference
between these two cases. While in the first one the
charged lepton plays only a ``spectator r{\^o}le'' and can be 
neglected to a good approximation, in the second case the two-particle
state has to be considered. 

Furthermore, we shall address the question whether charged leptons can
show an oscillation pattern. Usually it is argued that their
oscillation frequencies are too high to be observable and moreover,
that the coherence of wave packets corresponding to different mass
eigenstates is lost under experimental conditions. 

In general, an oscillation pattern
is observed if 1) the distance between the source and the 
detector is smaller than the coherence length $l_{\rm coh}$ and 
2) the size of the source and of the
detector are smaller than the oscillation length $l_{\rm osc}$.
In analogy to neutrino oscillations~\cite{coherence}, 
the coherence length for charged leptons can be written as 
\be
l_{\rm coh}= \sigma \; \frac{2E^2}{\Delta m^2} \,,
\ee
with the width of the wave packet $\sigma$, the energy
$E$ and mass difference $\Delta m^2=m^2_i - m^2_j$ of the charged leptons.
As an example we estimate $l_{\rm coh}$ for pion
decay in flight with an energy of $E_\pi\approx 40$~GeV.   
Using $\sigma \ap \gamma\tau $ and the pion half life 
$\tau = 2.6 \times 10^{-8}$~s one obtains   
$l_{\rm coh} = {\mathcal O}(10^8$~m).
The oscillation length $l_{\rm osc}=4\pi E/\Delta m^2$ would
amount to ${\mathcal O}(10^{-11}$~m). 
Consequently, the coherence condition could be met experimentally
while the oscillation pattern would be smeared out. If this were the 
complete argument, one still could hope to derive information about 
$V$ measuring only charged leptons. 
We will show however, that it is necessary to measure the accompanying 
neutrino simultaneously in order to obtain any information on $V$.

\section{Mixing in the lepton sector}

Let us first recall how fermion mass matrices are diagonalized. 
Generally, the mass terms in the Lagrangian are given by 
\be
 {\cal L}_{\rm mass} = - \sum_{\alpha,\beta} \bar \nu_{L,\alpha}
                       M_{\alpha\beta}^{(\nu)} \nu_{R,\beta}
                       - \sum_{\alpha,\beta} \bar l_{L,\alpha}
                       M_{\alpha\beta}^{({\rm ch})} l_{R,\beta} + {\rm h.c.},
\ee
and the mass matrices $M_{\alpha\beta}$ 
are not hermitian or even diagonal in the basis of the weak
eigenstates. (We denote the weak eigenstates
$\nu_\alpha=\{\nu_e,\nu_\mu,\nu_\tau\}$  and $l_\alpha=\{e,\mu,\tau\}$ 
by greek indices $\alpha,\beta\ldots$, 
and the mass eigenstates by latin indices $i,j,\ldots$ Furthermore,
we have assumed that the neutrinos have only Dirac mass terms to simplify
the formulas). Since the mass matrices are not hermitian, they cannot be
diagonalized by a simple unitary transformation. However, arbitrary
mass matrices can be diagonalized by a biunitary transformation~\cite{bi},
\be
 M_{\rm diag}^{(\nu)}=U^{(\nu)\dagger} M^{(\nu)} T^{(\nu)} \,,
\ee
where  $U^{(\nu)} U^{(\nu)\dagger} = T^{(\nu)} T^{(\nu)\dagger} ={\bf 1}$.
Then,  the connection between weak and mass eigenstates is given by
\be  \label{U}
 \nu_{R,\alpha} = \sum_i T^{(\nu)}_{\alpha i} \nu_{R,i} \:, \qquad\qquad
 \nu_{L,\alpha} = \sum_i U^{(\nu)}_{\alpha i} \nu_{L,i} \:, 
\ee
and similar equations hold for the charged leptons.
 
Inserting the transformations of Eq.~(\ref{U}) into the charged
current Lagrangian of the SM,
\be   
 \L_{CC} = -
 \frac{g}{\sqrt{2}}\sum_\alpha \bar l_{L,\alpha} \gamma^\mu \nu_{L,\alpha} 
                          W^-_\mu +{\rm h.c.} ,
\ee
results in
\be   \label{Hcc}
 \L_{CC} = -
 \frac{g}{\sqrt{2}}\sum_{i,j} \bar l_{L,i} \gamma^\mu V_{ij} \nu_{L,j} 
                              W^-_\mu +{\rm h.c.} , 
\ee
where we introduced the analogue of the CKM matrix in the lepton
sector \cite{MNS},
$V=U^{({\rm ch})\dagger}U^{(\nu)}$. Since the charged current
interaction involves only left-chiral fields of both charged leptons and 
neutrinos, the product of the two mixing matrices of the right-handed
leptons, $T^{({\rm ch})\dagger}T^{(\nu)}$, is unobservable.

In the case of massless neutrinos we can choose the neutrino mass
eigenstates arbitrarily. In particular, we can set
$U^{(\nu)}=U^{({\rm ch})}$ for any given $U^{({\rm ch})}$, hereby rotating
away the mixing. This shows that neutrino masses are a necessary
condition for non-trivial consequences of mixing in the lepton sector.

Finally, we recall that there are no flavor-changing neutral currents
within the standard model: The neutral current Lagrangian is diagonal
both in the weak eigenbasis,
\be   \label{Hnc}
 \L_{NC} = -
 \frac{g}{2\cos\theta_W}\sum_\alpha [\bar \nu_{L,\alpha} 
                                \gamma_\mu \nu_{L,\alpha}
 + \bar l_{\alpha} (g_V + g_A\gamma_5) \gamma_\mu l_{\alpha}] Z^\mu 
\ee
and in the mass basis
\be
 \L_{NC} = - 
 \frac{g}{2\cos\theta_W}\sum_i [\bar \nu_{L,i} \gamma_\mu \nu_{L,i} 
             + \bar l_{i}  (g_V + g_A\gamma_5) \gamma_\mu l_{i}] Z^\mu
\ee
due to the unitarity conditions,  $U^{(\nu)\dagger} U^{(\nu)} ={\bf 1}$ 
and $U^{({\rm ch})\dagger} U^{({\rm ch})} ={\bf 1}$.

\section{Pion decay and lepton mixing}

Many neutrino oscillations experiments use as source for the initial
lepton-neutrino state charged pions. In the SM without lepton mixing,
a tree-level calculation gives for the ratio $R$ of $\pi\to e\nu_e$ and
$\pi\to\mu\nu_\mu$ decay rates 
\be  \label{g1}
 R = 
 \frac{\Gamma(\pi\to e\nu_e)}{\Gamma(\pi\to\mu\nu_\mu)} =
 \frac{m_e^2}{m_\mu^2}\:\frac{(m_\pi^2-m_e^2)^2}{(m_\pi^2-m_\mu^2)^2} 
 \ap 1.28\times 10^{-4} \,.
\ee
Since angular momentum conservation  in the pion rest frame requires a
helicity flip of the lepton, the $S$-matrix elements of these decays
are proportional to the lepton masses $m_\alpha$ and, therefore, the
branching into electrons is suppressed. Hence, the two-particle state $\two$
created by a decaying positively charged pion is given by 
\be  \label{two}
 \two = \frac{1}{\sqrt{N}} \sum_{\alpha=e,\mu} m_\alpha 
        (1-m_\alpha^2/m_\pi^2) \: |l^+_\alpha \nu_\alpha \rangle \,,
\ee
where $N$ is a normalization constant. We have included the phase
space factor $1-m_\alpha^2/m_\pi^2$ into Eq.~(\ref{two})
because we assume that the state $\two$ lives for a macroscopic time
between its creation and detection. Therefore, both the lepton and the
neutrino are approximately on their mass-shell. 

Let us now examine what are the necessary changes if we want to account
for lepton mixing. Inserting Eq.~(\ref{U}) into  Eq.~(\ref{two})
we obtain 
\be   \label{twot}
 \twot = \frac{1}{\sqrt{N}} \sum_{i=1}^2 \sum_{j=1}^3 a_i V_{ij}
         \, |l_i^+ \nu_j \rangle \: e^{-\i (E_i+E_j)t} \,,              
\ee
where $a_i=  m_i(1-m_i^2/m_\pi^2)$. 

We are ordering the mass eigenstates according to the value of $m_i$,
i.e. $m_{l_1}<m_{l_2}<m_{l_3}$. Thus the state $l_3$ cannot be
populated in pion decay and, therefore, is omitted in the summation. 
Furthermore, we have assumed that all three neutrinos masses are
extremely small
compared to the electron mass as it is suggested by the currently favored
interpretation of neutrino oscillation experiments and cosmology.
Therefore, we could omit safely new terms in the $S$-matrix element
proportional to $m_{\nu_i}$, that in principle change the branching
ratio Eq.~(\ref{g1}) from its SM value~\cite{sh81}.
Note that the relative phases $V_{ij}$ of the different components 
$|l_i^+ \nu_j \rangle$ are fixed by
the Lagrangian, while the $a_i$ are real numbers. The time evolution
of $\twot$ is trivial, because we have expressed $\twot$ as a sum over
mass eigenstates.

In Eq.~(\ref{twot}), we have not displayed explicitly the finite
lifetimes $\tau=1/\Gamma$
of the states $l_2$ and $\nu_{2,3}$,  because this point is
not essential for our discussion. However, the finite lifetimes can
be restored treating the energy as a complex number,
$E=(m^2+p^2)-i\Gamma/2$ and noting that the decay products of $\two$
do not interfere with it.

Apart from the large difference between the lifetime of $l_2$ and of
$\nu_{2,3}$, there is another, more important, distinction between
neutrinos and charged leptons. If one decomposes $\two$ explicitly
into its basis states, then
\be   \label{deco}
 \two = \frac{1}{\sqrt{N}} \sum_{\alpha=1}^3 
        \left( \sum_{i=1}^2 a_i U^{({\rm ch})\ast}_{\alpha i}\: |l_i^+\rangle
         \right)  \otimes
        \left( \sum_{j=1}^3  U^{(\nu)}_{\alpha j} \: |\nu_j\rangle
         \right)  \,.
\ee
Defining a new basis appropriate for $\two$ by 
\be
 \two = \sum_{\alpha=1}^3 
        |l_\alpha^+\rangle_\pi \otimes |\nu_\alpha\rangle_\pi 
\ee
and comparing with Eq.~(\ref{deco}), it follows that the neutrino
state $|\nu_\alpha\rangle_\pi$ produced in pion decay is just a usual 
weak eigenstate,
$|\nu_\alpha\rangle_\pi = 
  \sum_{j=1}^3 U^{(\nu)}_{\alpha j}|\nu_{j}\rangle =
  |\nu_\alpha\rangle$.  
By contrast, the charged lepton state is
$|l^+_\alpha\rangle_\pi = 
 \sum_{i=1}^2 a_i/\sqrt{N} \: U^{({\rm ch})\ast}_{\alpha i}|l^+_i\rangle \neq 
 |l^+_\alpha\rangle$.
We will see below that it is the presence of the prefactors $a_1\neq a_2$
which allows the observation of neutrino oscillations in pion decay
without detecting the charged lepton. 
Note however also that $|l^+_\alpha\rangle_\pi \neq |l^+_\alpha\rangle$
even for $a_1=a_2$, because the component $l_3$ is missing.

Using $a_2\gg a_1$, we can approximate $\twot$ as
\be
 \twot \approx  \sum_{j=1}^3 
                 V_{2j} \: |l_2^+ \nu_j \rangle \: e^{-\i (E_2+E_j)t} 
\ee
with $a_2/\sqrt{N}\approx 1$. 
Clearly, one obtains in this approximation only neutrino oscillations,
because the charged lepton is in a pure mass eigenstate. Choosing
furthermore $V=U^{(\nu)}$, we obtain the state normally considered as 
initial state in pion decay,  
\be
 \two \ap  \sum_{\alpha=1}^3 \sum_{j=1}^3 
           \delta_{2\alpha} U_{\alpha j}^{(\nu)} \: |l^+_2 \nu_j \rangle 
      = |l^+_2 \nu_\mu \rangle \,.
\ee
The approximation $a_2\gg a_1$ which is widely used in textbooks is
numerically well justified. However, its use obscures the fact
that even for the choice $V=U^{(\nu)}$, i.e. identifying
mass and flavor eigenstates of the charged leptons, 
the charged lepton is nevertheless produced in a mixed state.

Let us now discuss different measurements of the exact two-particle
state, Eq.~(\ref{twot}). Since we are only interested in flavor
oscillations, we do not consider possible momentum measurements of the
two particles. Then, a measurement of the state $\twot$ is complete if
at time $t$ the quantum 
numbers $i$ or $\alpha$ of both the neutrino and the charged
lepton are determined. In the case that only one quantum number is observed,   
the probability $P(l)$ of this measurement is obtained by summing
over the quantum number of the unobserved particle, symbolically
$P(l)=\sum_{l'}P(l,l')$. 

\begin{figure}
\vskip1.5cm
\begin{picture}(0,0)%
\epsfig{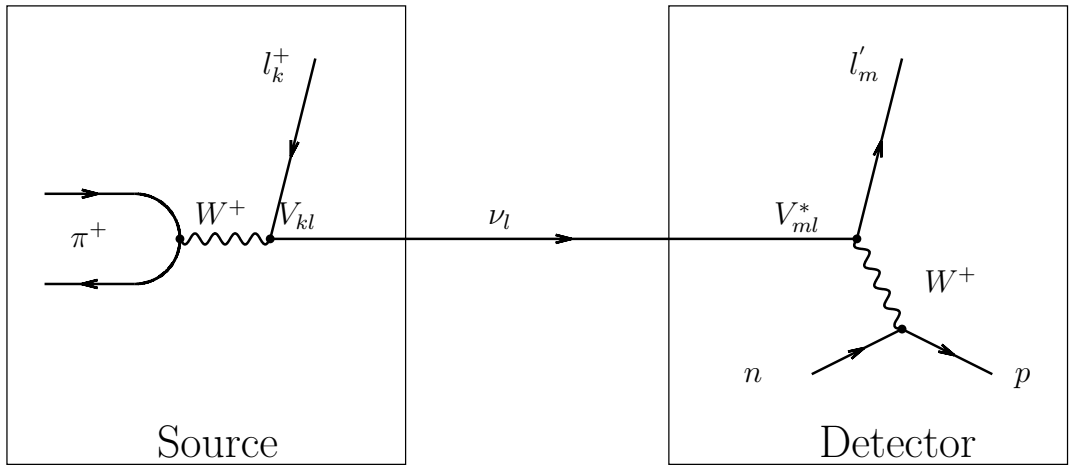}%
\end{picture}%
\setlength{\unitlength}{4144sp}%
\begingroup\makeatletter\ifx\SetFigFont\undefined%
\gdef\SetFigFont#1#2#3#4#5{%
  \reset@font\fontsize{#1}{#2pt}%
  \fontfamily{#3}\fontseries{#4}\fontshape{#5}%
  \selectfont}%
\fi\endgroup%
\begin{picture}(6369,2769)(304,-3223)
\put(811,-1906){\makebox(0,0)[b]{\smash{\SetFigFont{12}{14.4}{\rmdefault}{\mddefault}{\updefault}
$\pi^{+}$
}}}
\put(1216,-3166){\makebox(0,0)[lb]{\smash{\SetFigFont{17}{20.4}{\rmdefault}{\mddefault}{\updefault}
Source
}}}
\put(5176,-3166){\makebox(0,0)[lb]{\smash{\SetFigFont{17}{20.4}{\rmdefault}{\mddefault}{\updefault}
Detector
}}}
\put(1441,-1771){\makebox(0,0)[lb]{\smash{\SetFigFont{12}{14.4}{\rmdefault}{\mddefault}{\updefault}
$W^+$
}}}
\put(1936,-1771){\makebox(0,0)[lb]{\smash{\SetFigFont{12}{14.4}{\rmdefault}{\mddefault}{\updefault}
$V_{kl}$
}}}
\put(3196,-1771){\makebox(0,0)[lb]{\smash{\SetFigFont{12}{14.4}{\rmdefault}{\mddefault}{\updefault}
$\nu_l$
}}}
\put(1846,-871){\makebox(0,0)[lb]{\smash{\SetFigFont{12}{14.4}{\rmdefault}{\mddefault}{\updefault}
$l^{+}_{k}$
}}}
\put(5806,-2176){\makebox(0,0)[lb]{\smash{\SetFigFont{12}{14.4}{\rmdefault}{\mddefault}{\updefault}
$W^+$
}}}
\put(4726,-2716){\makebox(0,0)[lb]{\smash{\SetFigFont{12}{14.4}{\rmdefault}{\mddefault}{\updefault}
$n$
}}}
\put(6346,-2716){\makebox(0,0)[lb]{\smash{\SetFigFont{12}{14.4}{\rmdefault}{\mddefault}{\updefault}
$p$
}}}
\put(5356,-871){\makebox(0,0)[lb]{\smash{\SetFigFont{12}{14.4}{\rmdefault}{\mddefault}{\updefault}
$l^{'}_{m}$
}}}
\put(4906,-1771){\makebox(0,0)[lb]{\smash{\SetFigFont{12}{14.4}{\rmdefault}{\mddefault}{\updefault}
$V^{\ast}_{ml}$
}}}
\end{picture}
\vskip3cm
\caption{\label{pidecay}
Production of a superposition of neutrino mass eigenstates $\nu_l$ in
pion decay and subsequent detection of the neutrino flavour via the
secondary lepton $l^\prime_m$.}
\end{figure}

To begin with, we recall the
case normally treated in the literature, namely that the neutrino
flavor is detected while the lepton is not observed. In a first try,
we associate the probability 
$P(l_k,\nu_\alpha)=|\langle l_k \nu_\alpha \twot|^2$
to the measurement of the lepton mass eigenstate $k$ and the neutrino
flavor eigenstate $\alpha$ at time $t$. This would result in 
\be
 \langle l_k \nu_\alpha  \twot =  \frac{a_k}{\sqrt{N}}  \sum_{l=1}^3  
  V_{kl} U_{\alpha l}^{(\nu)\ast} e^{-\i (E_k+E_l) t} \,,
\ee
i.e. in an amplitude which does not only depend on $V$ but also on the
neutrino mixing matrix $U^{(\nu)}$. However, in practice one cannot
observe the flavor of a neutrino directly. Instead, the flavor of
the neutrino is determined looking at the mass eigenstates of the
charged lepton $l'$ produced in a secondary charged current reaction,
cf. Fig.~\ref{pidecay}.

Therefore, we should calculate
\be \label{yy}
 \langle l_k l_m^\prime| \hat H_{CC}(t) \twot =  \frac{a_k}{\sqrt{N}}
 \sum_{l=1}^3 V_{kl} V_{ml}^\ast \: e^{-\i (E_k+E_l)t} \,, 
\ee
where the action of $\hat H_{CC}$ destroys at time $t$ a neutrino
$\nu_\beta$ and creates a superposition of mass eigenstates of charged
leptons $l_m^\prime = U_{m\beta}^{({\rm ch})} l_\beta$. Here, $\hat
H_{CC}$ denotes the second quantized Hamiltonian of the usual
charged-current interaction. 
The corresponding probability to measure the primary lepton $l_k$ from the
pion decay and the secondary lepton $l_m^\prime$ produced by the neutrino is
\ba  \label{ww}
 P(l_k, l_m^\prime)  &=& \frac{a_k^2}{N} \Big\{
 \sum_{l=1}^3 |V_{kl}|^2 |V_{ml}|^2 +
 2 \sum_{n>l}^3 |V_{kl}V_{ml}^\ast V_{kn}^\ast V_{mn} | 
\nonumber\\  & &
 \,\cos\left[(E_l-E_n)t +\xi_{klmn}\right] \Big\} \,, 
\ea
where $\xi_{klmn}=\arg( V_{kl}V_{ml}^\ast V_{kn}^\ast V_{mn} )$.
If both charged leptons are observed, the probability (\ref{ww}) shows
clearly an oscillatory behavior.

In the case that only one of the two leptons is observed, the result
is completely different depending on if the primary $l_k$ or the
secondary $l_m^\prime$ (as indicator for the neutrino flavor) is
observed. In the first case, summing over $m$, we obtain
\be
 P(l_k)  =  \sum_m P(l_k, l_m^\prime)  = \frac{a_k^2}{N} 
\ee
using the unitarity of $V$, i.e. $\sum_k V_{ik}V_{jk}^\ast=\delta_{ij}$.
In the second case, we cannot
make use of these unitarity relations because the prefactors
$a_k$ depend on the summation index $k$. However, in the limit 
$a_2\gg a_1$, the result simplifies and we obtain the well-known
neutrino oscillation formula
\be
 P(l_m^\prime)  =  \sum_k P(l_k, l_m^\prime)  \ap  
 \left| \sum_l V_{2l} V_{ml}^\ast  
 \exp\left(-\i E_lt\right) \right|^2 \,.
\ee

Let us now comment shortly on the reason for this asymmetry. A flavor
oscillation experiment measures the correlation between the flavor or
mass quantum numbers of two particles. 
In the case of pion decay, we can use our knowledge about the
initial state created in the decay as replacement for an actual
measurement of the primary lepton. However, the kinematic of the decay
does not give us any useful information about the neutrino state: 
all three flavor eigenstates are with equal probability produced.
Therefore, the neutrino has to be measured via its secondary lepton to
obtain an oscillation pattern.

From this discussion, it should became clear that the difference
between charged leptons and neutrinos in $\two$ is rather specific to the
process considered: The V--A structure of the weak current together
with angular momentum conservation forbid the decay of the spinless
pion into two massless  spin 1/2 particles. Therefore, the pion decay
rate is proportional to the fermion masses $m^2$. Moreover, it is used
that the neutrino masses can be neglected compared to the masses of
the charged leptons.
Therefore, the fact that we have no a-priori information about the
neutrino but about the charged lepton is specific to pion decay.

Next, we want to discuss if it is possible to measure not the flavor
of a neutrino but its mass without measuring the lepton.    
A possible way to do this is to use Cherenkov or transition radiation
of the neutrino~\cite{rad}. In vacuum, neutrinos can interact with real
photons whose squared four-momentum $q^2$ is zero only via their
electromagnetic dipole or transition moments. In a medium, the photon
acquires a more complicated dispersion relation ($q^2\neq 0$) and, therefore,
neutrinos can emit real photons without decaying, $\nu_i\to\nu_i+\gamma$.
At least in principle, it is possible to reconstruct the mass of the
radiating neutrino from the spectra of emitted photons.

The probability to find the neutrino in the mass eigenstate $l$
without measuring the charged lepton is  
\be
 P(\nu_l) = \sum_{k=1}^2 P(l_k,\nu_l)
          = \frac{1}{N} \sum_{k=1}^2 a_k^2 |V_{kl}|^2 
          \ap \frac{a_2^2}{N} |V_{2l}|^2 \,.
\ee
In this case, we have the interesting result that $P(\nu_l)$
depends on the mixing matrix $V$ but does not show an oscillatory
behavior. Compared to a flavor measurement which introduces an
additional summation $\sum_{j}V_{jl}$  in the probability amplitude,
the phase of the probability amplitude of a mass measurement is constant 
and, consequently, no oscillation pattern can arise. 
On the other hand, $P(\nu_l)$ depends on $V$ because of the factors $a_k$.
Consequently, it is possible to extract information on the mixing
matrix $V$ measuring the mass eigenstate of the neutrino.

Finally, we comment briefly on charged lepton--neutrino states created
in nuclear beta decay. Due to the low nuclear energies involved, only the 
state $l_1$ is produced and, therefore, the charged lepton is
not only approximately but exactly in a pure state.

\section{$W$ decay and lepton mixing}

We discuss now the evolution of the two-particle state $|l\nu\rangle $ 
created by a decaying real $W$ boson. The $W$ boson is a spin-1 particle 
and, therefore, can decay into two massless fermions. Neglecting small
corrections of ${\cal O}(m^2/m_W^2)$, the state produced is equally
populated for all three generations. Thus, while the state 
\be  
 \twoW = \frac{1}{\sqrt{3}}\sum_{\alpha=1}^3 |l^+_\alpha \nu_\alpha \rangle 
\ee
is produced in the SM without lepton mixing by a decaying $W^+$, the state
\be  
 \twoW = \frac{1}{\sqrt{3}} \sum_{i,j=1}^3 V_{ij} |l_i^+ \nu_j \rangle 
\ee
is created with mixing.

We can repeat now the discussion of different measurements similar to the
case of $\two$. The only change necessary is the replacement of
$a_k/\sqrt{N}$ by $1/\sqrt{3}$. Hence, the probability to find the
primary lepton in a mass eigenstate $k$ and the secondary lepton in
$l_m^\prime$ becomes
\ba  \label{wwW}
 P(l_k, l_m^\prime)  &=& \frac{1}{3} \Big\{
 \sum_{l=1}^3 |V_{kl}|^2 |V_{ml}|^2 +
 2\sum_{n>l}^3 | V_{kl}V_{ml}^\ast V_{kn}^\ast V_{mn} |
\nonumber\\ & & 
  \cos\left[(E_l-E_n)t +\xi_{klmn}\right] \Big\} \,. 
\ea
In contrast to Eq.~(\ref{ww}), the probability is now symmetric  
in $l_k$ and $l_m^\prime$. In particular, the oscillation pattern
vanishes now in both cases as long as only one particle is observed.
Only when both the primary and the secondary lepton are observed,
an oscillation pattern according to Eq.~(\ref{wwW}) is observed.

We note that the same observation was made in Ref.~\cite{sm}. There,
the neutrino state produced in the decay of
a real $Z$ was examined. The authors of ~\cite{sm}
showed that also in this case neutrino flavor oscillations can be
observed, although the neutrinos are produced by neutral-current
interaction. Moreover, they showed that it is necessary to measure both
neutrinos in order to observe a oscillation pattern. Thus, their results
are in line with our findings presented above.

\section{Conclusion}

Flavor oscillations are observed by the detection of
correlations between two states. In an ideal experiment, the composition of
both states is measured. 
In experiments which use nuclear beta decay to produce the initial 
charged lepton--neutrino state the energy available is limited
to nuclear energies. Only the state $|l_1 \nu_j\rangle$ 
can be populated thus making a measurement of the charged lepton 
obsolete.

Experiments in which pion decay create
the initial lepton--neutrino state, one exploits the
known branching ratios into the different states $|l_i \nu_j\rangle$
as a substitute for the measurement of the charged lepton. 
These branching ratios differ only for different charged 
lepton mass eigenstates, but are the same for
different neutrino states. Therefore, the knowledge of the branching
ratios ``replaces'' only a measurement of the state of the charged
lepton, and the measurement of the neutrino state is necessary to
obtain information about lepton mixing.  
In contrast, the lepton--neutrino states produced in the decay
of real $W$ Bosons are symmetrical in their branching ratios. 
If experiments were carried out with such initial states,
both charged lepton and neutrino would be required to be measured 
in order to observe  flavor oscillations. 
In summary, the specific nature of the initial state used in oscillation
experiments explains the distinguished r{\^o}le of neutrinos compared
to charged leptons.


\widetext

\newpage

\end{document}